\documentclass[a4paper,11pt]{article}
\usepackage{jheppub} % for details on the use of the package, please see the JINST-author-manual
\usepackage[T2A]{fontenc}
\usepackage{tikz}
\usetikzlibrary{arrows.meta, positioning, quotes}
\usepackage{amsmath}
\usepackage{float}
\usepackage[normalem]{ulem}
\usepackage{graphicx}
\usepackage{multirow}
\def\bg{\bar{g}}
\def\bphi{\bar{\phi}}
\def\fzero{f_{(0)}}
\def\fone{f_{(1)}}
\def\ftwo{f_{(2)}}
\def\fthree{f_{(3)}}
\def\ffour{f_{(4)}}
\def\bT{\overline{T}}

\def\gzij{g_{(0)ij}}
\def\gtwoij{g_{(2)ij}}

\usepackage{color}

\title{\boldmath Holographic renormalization of Horndeski gravity}

\author[a,b]{Nicolás Cáceres}
\emailAdd{ntcaceres@uc.cl}
\affiliation[a]{Facultad de Física, Pontificia Universidad Catolica de Chile, Av. Vicu\~{n}a Mackenna 4860, Santiago, Chile }
\affiliation[b]{Departamento de Física, Universidad de Concepción, Casilla, 160-C, Concepción, Chile}

\author[c]{Crist\'obal Corral}
\emailAdd{cristobal.corral@uai.cl}
\affiliation[c]{Departamento de Ciencias, Facultad de Artes Liberales, Universidad Adolfo Ibáñez, Avda. Padre Hurtado 750, 2562340, Viña del Mar, Chile}

\author[d]{Felipe Díaz}
\emailAdd{f.diazmartinez@uandresbello.edu}
\affiliation[d]{Departamento de Ciencias Físicas, Universidad Andres Bello, Sazié 2212, Santiago de Chile}

\author[d]{Rodrigo Olea}
\emailAdd{rodrigo.olea@unab.cl}

\abstract{We study the renormalization of a particular sector of Horndeski theory. In particular, we focus on the nonminimal coupling of a scalar field to the Gauss-Bonnet term and its kinetic coupling to the Einstein tensor. Adopting a power expansion on the scalar function that couples the Gauss-Bonnet term, we find specific conditions on their coefficients such that the action and charges are finite. To accomplish the latter, we add a finite set of intrinsic boundary terms. The contribution of the nonminimal coupling generates an effective scalar mass, allowing us to recover a modified Breitenlohner-Freedman bound. Furthermore, we compute the holographic 1-point functions and Ward identities associated with the scalar field and the metric. We constrain the parameter space of the theory by taking into account the preservation of scaling symmetry at the boundary.  }

\begin{document}
\maketitle
\flushbottom

\section{Introduction}
Scalar fields play a fundamental role in the description of systems in areas as diverse as high-energy physics, gravity, cosmology, and condensed matter. They are essential to account for spontaneous symmetry breaking by acquiring a nonzero vacuum expectation value. This is the case of the behavior of superconductors but also of the mass-generating mechanism in the electroweak sector of the Standard Model of particle physics. 

In the more recent framework of anti-de Sitter/Conformal Field Theory (AdS/CFT) correspondence~\cite{Maldacena:1997re,Witten:1998qj,Gubser:1998bc,Aharony:1999ti}, coupling scalar fields to gravity introduces new sources and operators in the dual boundary CFT. Indeed, holographic superconductors~\cite{Hartnoll:2008vx,Albash:2008eh,Nishioka:2009zj,Montull:2009fe,Herzog:2010vz,Domenech:2010nf, Montull:2012fy, Salvio:2013jia,Cai:2015cya} and models featuring momentum dissipation~\cite{Andrade:2013gsa,Gouteraux:2014hca,Andrade:2016tbr} are examples of systems with physical properties captured by scalar fields. However, in black hole physics, different impersonations of the no-hair theorem pose strong limitations to finding nontrivial scalar configurations in stationary spacetimes~\cite{Bekenstein:1971hc,Bekenstein:1972ny,Bekenstein:1972ky,Sudarsky:1995zg,Bekenstein:1996pn}. Nevertheless, it is possible to circumvent these restrictions by introducing a cosmological constant and/or a nonminimal coupling between the scalars fields and the geometry. This allows to obtain black hole solutions in this class of scalar-tensor theories of gravity~\cite{Bocharova:1970skc,Bekenstein:1974sf,Martinez:1996gn,Martinez:1996uv,Martinez:2002ru,Martinez:2004nb,Babichev:2013cya} (for a review see~\cite{Barack:2018yly}). 

Finding sensible scalar-tensor theories whose dynamics and space of solutions are of relevance for holography is per se an interesting quest. However, one should start with those whose classical aspects have already been well understood, in order to find a reasonable dual CFT at the boundary. In these regards, a natural possibility is Horndeski gravity~\cite{Horndeski:1974wa}. This is constructed as the most general scalar-tensor theory that: (i) is invariant under diffeomorphisms, local Lorentz transformations and it is torsion-free,\footnote{For extensions of this theories in the presence of torsion see \cite{Barrientos:2017utp}, and \cite{Bahamonde:2022cmz, Armaleo:2023rhj} for further developments.} (ii) is constructed out of the metric, a scalar field and derivatives thereof, and (iii) possesses second-order field equations for the dynamical fields, avoiding the presence of ghosts. Their separated sectors have been studied in different contexts. In particular, the one that remains invariant under a constant shift of the scalar field has received a lot of attention in the last few years. This is usually referred to as the shift-symmetric sector of Horndeski theory. A specific term that belongs to the latter class is the nonminimal coupling between the Einstein tensor and the kinetic term of the scalar field. Indeed, when general relativity is augmented by this term, it has been shown that the space of solutions is endowed with black holes~\cite{Rinaldi:2012vy,Babichev:2013cya,Anabalon:2013oea,Cisterna:2014nua,Bravo-Gaete:2014haa,Babichev:2016rlq, Stetsko:2018fzt,Baggioli:2021ejg}, black strings~\cite{Cisterna:2018jsx,Cisterna:2018mww}, boson and neutron stars~\cite{Cisterna:2015yla,Brihaye:2016lin,Cisterna:2016vdx}, gravitational instantons~\cite{Bardoux:2013swa,Arratia:2020hoy}, among many other compact and extended objects (for a review see~\cite{Babichev:2016rlq}). 

In holography, models of momentum dissipation have been studied in shift-symmetric sectors of Horndeski gravity, allowing one to compute the DC conductivity analytically in the dual field theory~\cite{Jiang:2017imk}. Indeed, the same sector admits asymptotically Lifshitz black holes, providing a bridge for studying AdS/CMT correspondence~\cite{Bravo-Gaete:2013dca}. In the context of quantum information theory,  holographic entanglement entropy has been studied in Ref.~\cite{Caceres:2017lbr, DosSantos:2022exb} and, based on the complexity = action conjecture~\cite{Brown:2015bva}, the holographic complexity of AdS black holes with planar transverse sections have been computed in Ref.~\cite{Feng:2018sqm}. The thermodynamics of charged black holes in Horndeski gravity has been investigated in Ref.~\cite{Feng:2015oea}, showing that the usual viscosity bound can be violated depending on the mass/charge ratio. Thus, different holographic evidence has shown us that Horndeski gravity is a fruitful scenario for studying holography (for additional references, for instance, see~\cite{Kuang:2016edj,Jiang:2017imk,Filios:2018xvy,Li:2018kqp, Liu:2018hzo,Feng:2018sqm,Li:2018rgn,Santos:2021orr,Santos:2023flb,Santos:2023mee}).  

Another sector of Horndeski gravity that has drawn considerable attention in recent literature is the nonminimal coupling of scalar fields to the Gauss-Bonnet (GB) term. This interaction term appears naturally in the dimensional reduction of string-inspired models of gravity, e.g. Lovelock theory~\cite{Mueller-Hoissen:1987mas,Mueller-Hoissen:1987nvb,Mueller-Hoissen:1989bdj,Charmousis:2008kc,Charmousis:2012dw}, and different black hole and wormhole solutions have been found~\cite{Mignemi:1992nt,Kanti:1995vq,Torii:1996yi,Yunes:2011we,Sotiriou:2013qea,Sotiriou:2014pfa,Prabhu:2018aun,Antoniou:2019awm}. A remarkable feature of this interaction is that a scalar field coupled to the GB term can undergo a tachyonic instability, triggering a spontaneous scalarization of black holes~\cite{Doneva:2017bvd,Silva:2017uqg,Ripley:2019aqj,Silva:2020omi,East:2021bqk,Antoniou:2021zoy}. Indeed, it has been shown that such instability can end up with a scalarized black hole possessing either  rotation~\cite{Cunha:2019dwb,Berti:2020kgk,Hod:2020jjy,Elley:2022ept} or charge~\cite{Doneva:2018rou} (for a review see~\cite{Doneva:2022ewd}). From a holographic viewpoint, this phenomenon can be seen as a condensation process in the dual field theory, producing a critical phase transition characterized by a nonvanishing vacuum expectation value~\cite{Guo:2020sdu}. 

In view of possible holographic applications of Horndeski gravity, finding a suitable renormalization scheme in asymptotically AdS spacetimes for this theory has become a crucial point. In Einstein-AdS gravity, there exists an approach for rendering the Euclidean on-shell action and conserved charges finite in asymptotically AdS spacetimes, known as holographic renormalization. It is a scheme relevant in the context of gauge/gravity duality and it consists on the addition of intrinsic boundary counterterms~\cite{Balasubramanian:1999re,Chamblin:1999tk,Emparan:1999pm,deHaro:2000vlm,Skenderis:2002wp}. A different possibility, referred to as topological renormalization, consists in adding a single topological term to the bulk action, which acts as a counterterm, cancelling divergences in the action and its variation. This alternative has been widely explored in Refs.~\cite{Aros:1999id,Aros:1999kt,Olea:2005gb,Olea:2006vd,Miskovic:2009bm} and it has been proved to be equivalent to holographic renormalization for asymptotically AdS spaces with conformally flat boundaries~\cite{Anastasiou:2020zw}. For a general boundary geometry, topological renormalization needs to be supplemented by additional counterterms, which can be obtained by a proper embedding of Einstein gravity into conformal gravity as proposed in Ref.~\cite{Anastasiou:2020mik,Anastasiou:2021tlv,Anastasiou:2022ljq,Anastasiou:2023oro}, in an improved framework dubbed conformal renormalization (for the case including scalar couplings, see~\cite{Anastasiou:2022wjq}).

In Horndeski gravity, the variational principle with Dirichlet boundary conditions for the fields was studied in full detail in Ref.~\cite{Padilla:2012ze}. The holographic renormalization of a particular sector of the theory, i.e., the one considering the kinetic coupling to the Einstein tensor, was worked out in Ref.~\cite{Liu:2017kml}. Additionally, in Ref.~\cite{Feng:2015oea}, when computing the Euclidean on-shell action of the black hole in Ref.~\cite{Anabalon:2013oea},  it was pointed out that background subtraction methods do not match the result of holographic renormalization. Moreover, to first order in the saddle-point approximation, none of these approaches reproduce the computation of the black hole entropy via the Noether-Wald formalism~\cite{Wald:1993nt,Iyer:1994ys,Wald:1999wa}. Thus, a careful analysis of renormalization schemes in Horndeski gravity is a must in order to address this and other issues.

In this work, we analyze the renormalization of AAdS sector of Horndeski gravity. Taking the idea of topological renormalization as motivation, we consider an arbitrary coupling between the scalar field and the GB term. This is a suitable starting point since, for a constant scalar coupling, one recovers the standard results of holographic renormalization for pure gravity. In the general case, for a power expansion of the scalar coupling, we require specific conditions on their coefficients such that the action and charges are finite. In the presence of a quadratic sGB (sGB) coupling, a modified Breitenlohner-Freedman (BF) bound is obtained by demanding that the holographic stress-energy tensor remains traceless. These conditions provide a unitary CFT with scalar operators at the boundary.

This article is organized as follows. In Sec.~\ref{Sec:HoloReno}, we shortly review the holographic renormalization procedure of four-dimensional Einstein gravity minimally coupled to a scalar field and fix notation. In Sec.~\ref{Sec:SGB}, we move forward and consider Einstein sGB gravity. We show that an asymptotic analysis restricts the possible couplings in order to have a consistent holographic theory. In Sec.~\ref{Sec:Horni}, we obtain the proper boundary counterterms in the presence of the scalar-kinetic coupling to the Einstein tensor and we show how the BF bound is modified in this case. Section~\ref{Sec:Concs} is devoted to summarizing our results. Additionally, we include Appendix~\ref{App:BdryConditions} for details about the boundary conditions for the scalar field.

\section{Holographic renormalization with minimally-coupled scalar fields}\label{Sec:HoloReno}
%%%%%%%%%%%%%%%%%%%%%%%%%%%%%%%%%
Anti-de Sitter is a maximally symmetric space with negative constant curvature, whose conformal compactification results in a manifold with a conformal boundary. 
The line element of AdS$_{d+1}$ spacetime can be expressed in Poincaré coordinates as
\begin{align}
    ds^2={\tilde g}_{\mu\nu}dx^\mu dx^\nu = \frac{\ell^2}{z^2}\left(dz^2 + \eta_{ij}dx^i dx^j\right)~,
\end{align}
where  $x^i = (t, x_1, \dots, x_{d-1})$ are boundary coordinates, $\eta_{ij}$ represents the Minkowski metric, $z$ is the coordinate along the additional bulk dimension, and $\ell$ stands for the AdS radius, which is related to the cosmological constant by $\Lambda = -d(d-1)/2\ell^2$. 

In order to overcome the boundary singularity, one can employ a conformal compactification of the spacetime through a Weyl rescaling denoted by $g = z^2{\tilde g}$, where the conformal boundary is located at $z=0$. As a result, the conformal metric $g$ smoothly extends to the boundary and defines the boundary metric as
\begin{align}
    g_{(0)} = \lim_{z\to 0}z^2{\tilde g}~.
\end{align}
The same procedure can be extended for more general spaces \cite{Penrose:1985bww}. In particular, asymptotically AdS (AAdS) spacetimes are defined in this way, such that
the bulk metric is written as 
\begin{align}
    ds^2 = \frac{\ell^2}{z^2}\left(dz^2 + {\bg}
    _{ij}(z,x^i)dx^i dx^j\right)~.
\end{align}
By construction, $\bg_{ij}$ has a smooth limit as $z\to0$, such that it admits the Fefferman-Graham (FG) expansion ~\cite{Fefferman:1984asd}
\begin{align}
    \bg_{ij} = g_{(0)ij} + z g_{(1)ij} + z^2 g_{(2)ij}+\dots~.
\end{align}
The coefficients $g_{(n)ij}$ with $n\neq0$ can be obtained by solving Einstein's equations order by order. In doing so, the coefficients of the odd powers of $z$ are set to zero. Here, for later convenience, we will choose the radial coordinate as $\rho = z^2$. Then, the metric for AAdS spaces looks like
\begin{align}
    ds^2 = \frac{\ell^2d\rho^2}{4\rho^2} + \frac{\ell^2}{\rho}\bg_{ij}dx^i dx^j~,
\end{align}
where the asymptotic expansion of the boundary metric is given by
\begin{align}
    \bg_{ij} = g_{(0)ij} + \rho g_{(2)ij} + \dots + \rho^{d/2}\left(g_{(d)ij} + h_{(d)ij}\log\rho \right)+\dots~.
\end{align}
The coefficient $h_{(d)ij}$ only appears in even boundary dimensions and the ellipsis denotes higher powers of $\rho$ which correspond to non-normalizable modes.  

The near-boundary form of the metric realizes the asymptotically AdS condition on the Riemann tensor 
\begin{align}
    R_{\mu\nu\lambda\sigma} = \frac{1}{\ell^2}\left({\tilde g}_{\mu\lambda}{\tilde g}_{\nu\sigma} - {\tilde g}_{\mu\sigma}{\tilde g}_{\nu\lambda}\right) + {\cal O}\left(\rho\right)~.
\end{align}
The FG expansion unveils the infrared divergences in the gravity action.\footnote{In the AdS/CFT correspondence, the infrared behavior of the gravitational theory is mapped to the physical properties of the gauge theory in the ultraviolet \cite{Susskind:1998dq}.} In order to remove these divergences, one can introduce counterterms, which are the result of solving the Einstein equations to determine the expansion coefficients $g_{(n)ij}$, in terms of $g_{(0)}$.
 Finally, one inverts the relations to obtain $g_{(n)ij}$ as intrinsic covariant quantities from the point of view of the boundary metric, such that divergences of the action are removed. This procedure is referred to as \emph{holographic renormalization} and it was proposed in Refs.~\cite{Henningson:1998gx, deHaro:2000vlm}. It is a systematic method for extracting holographic quantities such as correlation functions, Ward identities, and Weyl anomalies (see \cite{Skenderis:2002wp} for a review).

In a similar way that the boundary of AdS defines a conformal class of metrics, a bulk field does not induce a specific one on the boundary. Thus, when imposing boundary conditions in AdS, they have to deal with a conformal class rather than any specific representative, a key ingredient to set the boundary value problem in AdS/CFT duality. For the metric tensor, it is imperative to take a Dirichlet condition on $g_{(0)ij}$ instead of fixing the boundary metric $h_{ij}=\bg_{ij}/\rho$; the latter obviously exhibits a divergent behavior. Therefore, the role of the counterterms is two-fold: (i) they ensure the finiteness of the on-shell action, and (ii) they define a well-posed variational principle for $g_{(0)ij}$.

With a renormalized action, one obtains the holographic stress tensor by performing a variation with respect to the boundary metric. The Gubser-Klebanov-Polyakov-Witten dictionary relates the low-energy limit of String Theory (semiclassical regime of supergravity on an AAdS background) to the generating functional of a gauge theory with conformal symmetry at its boundary. The duality can be stated as 
\begin{align}
{\rm exp}\left(i S_{\rm grav}[\phi_{(0)}^I]\right) = \left\langle {\rm exp}\left(i\int_{\partial\cal M}d^dx~ \phi_{(0)}^I {\cal O}_I\right) \right\rangle~,
\end{align}
where $S_{\rm grav}$ is a gravitational action with dynamical fields living on an AAdS$_{d+1}$ background, $\phi_{(0)}^I$ is the value of the fields at the conformal boundary, and ${\cal O}_I$ is the collection of gauge operators sourced by $\phi_{(0)}^I$. Then, a gravitational action on a $(d+1)$-dimensional AAdS spacetime is related to the quantum effective action of a $d$-dimensional CFT. Considering a scalar field in an Euclidean AdS gravity, the quantum generating functional of the dual CFT living on a background metric $g_{(0)}$ reads
\begin{align}
    Z_{\rm CFT}[g_{(0)},\phi_{(0)}] = \left\langle{\rm exp}\left[\int_{\partial\cal M}d^dx\sqrt{g_{(0)}}\left(\frac12g_{(0)ij}T^{ij}+\phi_{(0)}{\cal O}\right)\right]\right\rangle~,
\end{align}
where $T^{ij}$ is the CFT stress-energy tensor and ${\cal O}$ is a scalar operator sourced by $\phi_{(0)}$. Then, one obtains a bulk/boundary relation, that is,
\begin{align}\label{Tholog}
    \langle T^{ij}\rangle = \frac{2}{\sqrt{-g_{(0)}}}\frac{\delta S_{\rm ren}}{\delta g_{(0)ij}} =  \lim_{\rho\to0}\frac{2}{\sqrt{-\bg}}\frac{\delta S_{\rm ren}}{\delta \bg_{ij}} = \lim_{\rho\to0}\left(\frac{1}{\rho^{d/2-1}}T^{ij}[h]\right)~,
\end{align}
where $T^{ij}[h]$ is the stress tensor of the renormalized action.  This tensor is made of two parts: the first one is the canonical momentum which comes from the addition of the Gibbons-Hawking-York (GHY) term to the Einstein-Hilbert action ~\cite{Brown:1992br}. The second contribution arises from the variation of the counterterms introduced to renormalize the theory.
The resulting holographic stress-energy tensor is, indeed, finite and it corresponds to a boundary operator that is dual to the bulk gravitational field. This mapping between the asymptotic behavior of a bulk field and the source of a boundary quantum operator in the dual CFT is a fundamental aspect of the holographic dictionary. 

For matter fields coupled to gravity, the procedure is analog. It involves the asymptotic expansion of bulk fields near the conformal boundary. %The leading-order behavior corresponds to the Dirichlet data that serves as the source for the dual operator in the holographic theory. The subleading coefficients, on the other hand, can be determined by solving the bulk field equations, providing the necessary information to identify the divergent terms in the action. Then, one can construct intrinsic covariant quantities that cancel out these divergences.
In a general setting, the 1-point function of any field appears as the undetermined, subleading term as described in Ref.~\cite{Skenderis:2002wp}. The coefficient is determined by the boundary conditions imposed on the field, allowing for a comprehensive understanding of the field's behavior and its connection to the boundary theory. Here, we focus on scalar operators that exhibit nontrivial interactions with the boundary metric.

Let us examine first the four-dimensional Einstein-AdS gravity minimally coupled to a massive scalar field. The action of such a theory is given by
\begin{align}\label{Imin}
    S_{\rm min} = \frac12\int_{\mathcal M}d^4x \sqrt{-g}\left(\frac{R+6\ell^{-2}}{\kappa} - \left(\partial\phi\right)^2 -m^2\phi^2\right) -\frac{1}{\kappa}\int_{\partial{\cal M}}d^3x\sqrt{-h}\,K~,
\end{align}
where ${\cal M}$ is an AAdS$_4$ manifold, $\kappa$ is the gravitational constant related to the Newton's constant by $\kappa=8\pi G$, $R=g^{\mu\nu}R_{\mu\nu}$ is the Ricci scalar, $\ell$ is the AdS radius, $m$ is the mass parameter of the scalar field $\phi$, and $K$ is the trace of the extrinsic curvature of the boundary surface, denoted by ${\partial\cal M}$. An arbitrary variation of the action gives 
\begin{align}
    \delta S_{\rm min} ={}&\int_{\cal M}d^4x\sqrt{-g}\left( \frac{1}{2\kappa}\left(E_{\mu\nu} - \kappa T_{\mu\nu}\right)\delta g^{\mu\nu} + \left(\Box\phi - m^2\phi\right)\delta\phi\right)\nonumber \\ {}& - \int_{\partial\cal M}d^3x\sqrt{-h}\left(\pi_{ij}\delta h^{ij} + \pi_\phi \delta\phi\right)~,
\end{align}
where, in the bulk, we have defined
\begin{align}
    E_{\mu\nu} ={}& R_{\mu\nu}-\frac12 R g_{\mu\nu} + \Lambda g_{\mu\nu}~,\nonumber \\ 
    T_{\mu\nu} ={}& \nabla_{\mu}\phi\nabla_{\nu}\phi - \frac12 g_{\mu\nu}\left(\nabla^\lambda \phi \nabla_\lambda \phi + m^2 \phi^2\right) ~,\label{Tmunuminimalcoupling}
\end{align}
and the boundary variations define the canonical momenta associated with the radial evolution of the metric and the scalar field by
\begin{align}
    \pi_{ij} = \frac{1}{2\kappa}\left(K_{ij} - K h_{ij}\right) \;\;\;\;\; \mbox{and} \;\;\;\;\; \pi_\phi = n^\mu\partial_\mu \phi ~,
\end{align}
respectively. Thus, the field equations for the metric and the scalar field can be read as
\begin{align}\label{EOMmin}
   E_{\mu\nu} - \kappa T_{\mu\nu} {}&=0~, \nonumber \\ \Box\phi -m^2\phi {}&=0~.
\end{align}

In what follows, we shall consider the FG expansion with no logarithmic modes, which describes a wide class of gravitational setups. In that case, the scalar field can be expanded asymptotically as
\begin{align}\label{phiminexp}
    \phi(\rho,x^i) = \rho^{(3-\Delta)/2}\bar{\phi} = \rho^{(3-\Delta)/2}\left( \phi_{(0)} + \rho\phi_{(2)}+\dots+\rho^{(2\Delta-3)/2}\phi_{(2\Delta-3)}+\dots \right)~,
\end{align}
where $\Delta$ is a constant to be determined. The holographic renormalization method for the three-dimensional version of Eq.\eqref{Imin} with $\Delta = 2$ was studied in Ref.~\cite{deHaro:2000vlm}. Inserting the asymptotic expansion of the fields into the equation of motion for the metric leads to the following system of equations
\begin{align}\notag
 0 &=  {\cal R}_{ij}(\bg) - \partial_\rho \bg_{ij} - \bg_{ij}\bg^{mn}\partial_\rho \bg_{mn} + {}\rho\left(2\bg_{im}\partial_\rho\left(\bg^{mn}\partial_{\rho}\bg_{nj}\right)+\bg^{mn}\partial_\rho\bg_{mn}\partial_\rho\bg_{ij}\right) \\ &+{\kappa}\rho^{2-\Delta}\left(\tfrac{1}{2}m^2\ell^2\bphi^2\bg_{ij} + \rho\partial_i\bphi\partial_j\bphi\right)~, \nonumber
\\ \label{ExpnEOMmin} 0 &= \bg^{mn}\partial_\rho \bg_{ml}\bg^{lp}\partial_\rho\bg_{pn} + 2\partial_\rho(\bg^{mn}\partial_{\rho}\bg_{mn}) \\  &+\kappa\rho^{1-\Delta}\left(\left(\tfrac12 m^2\ell^2+(3-\Delta)\right)\bphi^2+4\left((3-\Delta)\rho +\rho^2\right)\bphi\partial_\rho\bphi  \right)\notag \\ 0&=\nabla_i\left(\bg^{jm}\partial_\rho\bg_{jm} \right)-\nabla_j\left(\bg^{jm}\partial_\rho\bg_{mi} \right) + \kappa\rho^{2-\Delta}\left((3-\Delta)\bphi\partial_i\bphi + 2\rho\partial_\rho\bphi\partial_i\bphi\right)~,\nonumber
\end{align}
where ${\cal R}_{ij}(\bg)$ is the Ricci tensor of the metric $\bg$. The equation of motion for the scalar field, in turn, can be expressed as
\begin{align}\label{ExpnEOMminscalar}
    0={}&\left(\Delta(\Delta-3)-m^2\ell^2\right)\bphi + \rho\left(\bg_{ij}\partial_i\partial_j\bphi - 2(5-2\Delta)\partial_\rho\bphi +(3-\Delta)\partial_\rho(\log\bg)\bphi \right)\nonumber \\ {}&+\rho^2\left(2\partial_\rho(\log\bg)\partial_\rho\bphi + \partial_\rho^2\bphi\right)~.
\end{align}
At zeroth order in the holographic coordinate, this equation gives rise to a relation between the mass and $\Delta$ given by
\begin{align}
    m^2\ell^2 = \Delta(\Delta -3)~.
\end{align}
As shown in Refs.~\cite{Gubser:1998bc, Witten:1998qj}, this relation corresponds to the one between the mass of a scalar field on an AdS background and the conformal dimension of the dual operator at the boundary. The latter is determined by the rescaling properties of a scalar operator in the CFT and it can be obtained by analyzing the 1-point function of the holographic operator. It is worth noticing that, in a unitary dual theory, the scaling dimension of a scalar operator must be a positive integer, which defines constraints on the allowed values of the mass of the corresponding bulk field. The BF bound~\cite{Breitenlohner:1982bm, Klebanov:1999tb}, further restrict the mass to satisfy
\begin{align}\label{BFbound}
  -\left(\frac{d}{2}\right)^2 < m^2\ell^2 ~.
\end{align}
This bound serves to understand the interplay between bulk and boundary physics in AdS/CFT correspondence, matching the stability of the bulk scalar with the unitarity of the dual theory. Indeed, an upper bound on the scalar mass can be obtained if alternate quantization is allowed. The latter appears if one imposes that the scaling dimension in the alternate quantization scheme is above the unitarity bound. Nevertheless, here we focus on the standard quantization scheme that restricts the scalar mass according to Eq.~\eqref{BFbound}.

To determine the coefficients in the expansion~\eqref{phiminexp}, one needs to solve the field equations~\eqref{ExpnEOMminscalar} order by order in the holographic coordinate. This process, however, cannot be carried out for a generic $\Delta$ and it needs to be addressed case by case, as discussed in Ref.~\cite{deHaro:2000vlm}.
Looking at the scalar field expansion in Eq.~\eqref{ExpnEOMminscalar}, at first-order in the holographic coordinate, there appears a second-order differential equation for the scalar field which also involves first derivatives of the boundary metric. In turn, in the equations of motion for the metric, the lowest-order term in $\rho$ is proportional to $\rho^{2-\Delta}$. To ensure a consistent asymptotic expansion, one should demand that $\Delta \leq 3$. For instance, when $\Delta = 2$, a self-interaction emerges between the boundary fields at the leading order, producing a backreaction on the geometry of the dual CFT; this represents a critical value of the conformal dimension. Alternatively, for $\Delta = 1$, the kinetic term backreacts on the boundary only at the next-to-leading order. Similarly, when $\Delta = 0$, the bulk scalar field becomes massless and the backreaction mainly comes from the kinetic term of the boundary scalar. These different choices of $\Delta$ lead to distinct behaviors and interactions between the bulk and boundary fields, offering valuable insights into the AdS/CFT correspondence. 

Let us focus on the case $\Delta=2$ or, equivalently, $m^2\ell^2 = -2$. This value is, indeed, admissible by the BF bound. Then, we proceed to solve the Einstein equations to determine the first coefficient in the metric expansion in terms of the sources. The coefficient reads
\begin{align}
    g_{(2)ij} = -{\cal S}_{ij}(g_{(0)}) - \frac{\kappa}{4}\phi^2_{(0)}g_{(0)ij}~,
\end{align}
where 
\begin{align}
    S_{ij}(g_{(0)}) := {\cal R}_{ij}(g_{(0)})-\frac14 g_{(0)ij}{\cal R}(g_{(0)})~,
\end{align}
is the Schouten tensor of the boundary metric $g_{(0)}$. An arbitrary variation of the on-shell action and using the asymptotic expansion yields
\begin{align}
    \delta S_{\rm min} ={}& \frac{\ell^2}{2\kappa}\int_{\partial\cal M} d^3x \sqrt{-g_{(0)}}\rho^{-1/2}\Bigg[\left(-\frac{2}{\rho}g_{(0)ij}-3g_{(2)ij} + {\cal O}\left(\rho^{1/2}\right)  \right)\delta g_{(0)}^{ij}\nonumber \\  {}&-\ell^2\left(\phi_{(0)} + {\cal O}\left(\rho^{1/2}\right)\right)\delta\phi_{(0)}\Bigg]~.
\end{align}
Therefore, in order to preserve a well-posed variational principle, one needs to add only intrinsic boundary terms. Indeed, inverting the series, one finds that the surface terms needed are
\begin{align}
    S_{\rm ct} + S'_{\phi} = 
    \frac{1}{\kappa}\int_{\partial \cal M}d^3x \sqrt{-h}\left(\frac{2}{\ell} +\frac{\ell}{2}{\cal R}(h) \right) + \frac{1}{\kappa}\int_{\cal M}d^3x \sqrt{-h}\left(\frac{\phi^2}{2\ell} \right),
\end{align}
which renormalize the gravity sector~\cite{Balasubramanian:1999re,deHaro:2000vlm} and the counterterm for a massive scalar field on AdS which cures the divergences coming from the scalar sector~\cite{Klebanov:1999tb}.\footnote{The covariant counterterms for the massive scalar including logarithmic modes have been found explicitly in Ref.~\cite{deHaro:2000vlm}, up to second order.} Then, the action
\begin{align}\label{Iminren}
    S_{\rm min}^{\rm ren} = S_{\rm min} + S_{\rm ct}+ S'_{\phi}~,
\end{align}
is finite on shell. Nonetheless, it is possible to add an extra term 
\begin{align}
    S'_{\partial\phi} = \frac{\gamma}{2\kappa}\int_{\partial\cal M}d^3x \sqrt{-h}\;\phi\,n^\mu\partial_\mu\phi~,
\end{align}
where $n^\mu$ is the outward-pointing unit normal to the boundary. Its coupling $\gamma$, by the use of a Legendre transformation, redefines the mass of the scalar field. This extra boundary term has been considered in Refs.~\cite{Lu:2013ura, Lu:2014maa} to obtain the correct thermodynamics for a given $\gamma$ which matches the ADM mass. At first glance, this term may be at odds with a variational principle based on mixed boundary conditions (see Appendix~\ref{App:BdryConditions}). However, within a holographic framework, what is relevant is that the variation of the action is finite and written down in terms of the variation of the holographic sources \footnote{A similar discussion, for the metric field, leads to extrinsic counterterms in AdS gravity \cite{Anastasiou:2020zw}.} That is the reason why, one can replace this extrinsic term by another which depends explicitly on the sources and the boundary conditions~\cite{Anabalon:2015xvl,Caldarelli:2016nni}.
Then, the counterterm for the scalar field is
\begin{align}\label{Sphi}
    S_\phi = \int_{\partial\cal M}d^3x\sqrt{-h}\left(\frac{\phi^2}{2\ell} + \frac{W\left(\phi_{(0)}\right)}{\ell\phi_{(0)}^3}\phi^3\right)~,
\end{align}
where $W\left(\phi_{(0)}\right)$ is determined by the boundary conditions (see Appendix~\ref{App:BdryConditions}) such that
\begin{align}\label{dWJ}
    \phi_{(1)} = \frac{dW}{d\phi_{(0)}}~.
\end{align}
Considering these new counterterms is crucial for establishing a well-defined variational principle for the source of the dual scalar operator under various boundary conditions. In the presence of mixed boundary conditions, the additional contribution from Eq.~\eqref{Sphi} introduces finite terms that represent multi-trace deformations of the CFT generating functional, potentially disrupting scaling symmetry depending on the conformal weight of the scalar operator. In the case of $\Delta = d$, the deformations are marginal, and conformal symmetry is preserved. Hence, the mass of the scalar field is important for a well-defined holographic CFT. Later, we illustrate how coupling the GB invariant through a scalar field function introduces an effective mass for the bulk scalar. This effective mass is constrained by the unitary bound and, under certain boundary conditions, by scaling symmetry. The latter constrains the parameter space of the scalar-tensor theory, crucial in the effective description of string theory beyond the leading order in the string coupling~\cite{Callan:1985ia,Gross:1986mw}.

The renormalized Euclidean action for the minimally coupled scalar with $\Delta = 2$ reads
\begin{align}
    S_{\rm min}^{\rm ren} %={}&  S_{\rm min} + S_{\rm ct} + S_{\phi} \nonumber \\ 
    ={}&   -\frac{1}{2}\int_{\cal M}d^4x \sqrt{g}\left(\frac{R+6\ell^{-2}}{\kappa} - \left(\partial\phi\right)^2 - m^2\phi^2\right) \nonumber \\ {}&+\frac{1}{\kappa}\int_{\partial\cal M}d^3x\sqrt{h}\left[K - \frac{2}{\ell} - \frac{\ell}{2}{\cal R}(h) - \kappa\left(\frac{\phi^2}{2\ell} + \frac{W\left(\phi_{(0)}\right)}{\ell\phi_{(0)}^3}\phi^3\right)\right]~.
\end{align}
Additionally, the quasi-local stress-energy tensor is given by 
\begin{align}
    T_{ij}[h] = -\frac{1}{\kappa}\left[K_{ij}-K h_{ij} + \frac{2}{\ell}h_{ij}-\ell\left({\cal R}_{ij} - \frac12 {\cal R}h_{ij}\right)\right] - h_{ij}\left(\frac{\phi^2}{2\ell} + \frac{W\left(\phi_{(0)}\right)}{\ell\phi_{(0)}^3}\phi^3\right)~,
\end{align}
and it provides a regular holographic stress tensor through Eq.~\eqref{Tholog}. In the next section, we follow to same procedure to find the covariant counterterms in theories of gravity with a non-minimally coupled scalar field.  
%%%%%%%%%%%%%%%%%%%%%%%%%%%%%%%
%%%%%%%%%%%%%%%%%%%%%%%%%%%%%%%

\section{Scalar-Gauss-Bonnet gravity}\label{Sec:SGB}
Gravitational dynamics arising from the low-energy limit of string theory is not governed solely by the Einstein equations. Indeed, higher-derivative terms induce effective theories with nontrivial interactions between the dilaton and the gravitational field. In the case of the heterotic string, the effective field theory contains higher-derivative terms at the first order in $\alpha'$ corrections~\cite{Callan:1985ia, Gross:1986mw} and its dynamics is dictated by the scalar-GB (sGB) action~\cite{Boulware:1986dr,Cano:2021rey}.
Although these modifications are small compared with the string coupling from an effective field theory viewpoint, they can be used to study scalar hair condensation holographically with/without spontaneous symmetry breaking~\cite{Guo:2020sdu}, as well as excitations of the ground state of superconductors~\cite{Bao:2021wfu}. 
In the gauge/gravity correspondence, $\alpha'$ corrections correspond to the t'Hooft coupling corrections in the dual field theory. In this section, we study how these stringy-generated gravity theories would modify the holographic correlators of the scalar and graviton fields.

Let us consider sGB. This theory represents a particular sector of Horndeski gravity and it involves the coupling of the GB invariant with an arbitrary smooth function of the scalar field, building on top of the action in Eq.~\eqref{Imin} while omitting the GHY term. Specifically, the action for sGB gravity is 
\begin{align}\label{GB}
    S_{\rm sGB} = \frac12\int_{\mathcal M}d^4x \sqrt{-g}\left(\frac{R+6\ell^{-2}}{\kappa} - \left(\partial\phi\right)^2 - m^2\phi^2 +2f(\phi){\cal G}\right)~,
\end{align}
where the GB term is given by
\begin{align}
    {\cal G} \equiv R_{\mu\nu\lambda\sigma}R^{\mu\nu\lambda\sigma} - 4R_{\mu\nu}R^{\mu\nu} + R^2~.
\end{align}
The field equations of the theory from arbitrary variations with respect to the metric and the scalar field, which yield
\begin{align}
    R_{\mu\nu}-\frac12g_{\mu\nu}R+\Lambda g_{\mu\nu} ={}& \kappa\left(T_{\mu\nu} + C_{\mu\nu}\right)~, \nonumber \\ \Box\phi -m^2\phi={}& -f'(\phi)\left(R_{\mu\nu\lambda\sigma}R^{\mu\nu\lambda\sigma}- 4R_{\mu\nu}R^{\mu\nu} +R^2\right)~,
\end{align}
respectively, where $T_{\mu\nu}$ has been defined in Eq.~\eqref{Tmunuminimalcoupling} and
\begin{align}\label{Cmunu}
    C^{\mu}_{\nu} ={}& -2\delta^{\mu\alpha\beta\gamma}_{\nu\lambda\sigma\tau}R_{\beta\gamma}^{\sigma\tau}\nabla^{\lambda}\nabla_{\alpha}f(\phi)\,.
\end{align}
Let us consider a power-series expansion of the scalar coupling function
\begin{align}\label{fphi}
    f(\phi) = \sum_{n=0}^{\infty}f_{(n)}\phi^n~.
\end{align}
Using the same asymptotic expansion as in the minimally coupled scalar [cf. Eq.~\eqref{phiminexp}], the scalar equation becomes
\begin{align}
24f_{(1)}\rho^{(\Delta-3)/2}+\left(48f_{(2)}+\ell^2\left(\Delta(\Delta-3)-\ell^2 m^2\right)\right)\bphi+\mathcal{O}\left(\rho^{(\Delta+1)/2}\right) = 0~.
\end{align}
From this expression, one can see that the quadratic sGB coupling contributes to an effective mass for the scalar field, say $m_{\rm eff}$, that is defined through the relation
\begin{align}\label{meff}
    m_{\rm eff}^2\ell^2 := m^2\ell^2 - \frac{48f_{(2)}}{\ell^2}\,.
\end{align}

The explicit form of the field equations turns out to be lengthy and we shall not present them here. However, the leading-order analysis indicates that $\Delta>1$ is necessary to have a nontrivial scalar source. Additionally, if $\Delta > 3$, there is no backreaction of the scalar field on the boundary metric, giving a trivial scalar source. Thus, similar to the minimally coupled case, we select $\Delta = 2$ for our analysis. This choice leads to more interesting dynamics, including the interaction between the scalar field and the boundary metric. Notice that, for this value of $\Delta$, the scalar field gives a finite contribution to the on-shell action at the cubic order. 
Then, the relation between $\Delta$ and the mass of the scalar field becomes
\begin{align}
    m^2\ell^2 - \frac{48}{\ell^2}f_{(2)} \equiv m_{\rm eff}^2\ell^2 = \Delta(\Delta-3)  = -2~.
\end{align}
Then, in order to have a well-defined unitary quantum field theory at the boundary, the effective mass of the scalar field must satisfy the BF bound, that is,
\begin{align}\label{BFmeff}
    -\frac{9}{4} < m_{\rm eff}^2\ell^2~.
\end{align}

The asymptotic analysis of Einstein equations requires that $\fone = 0$. Furthermore, the consistency of higher-order terms in the expansion requires that either $\fthree = 0$ or that the boundary value of the scalar field satisfies $\phi_{(0)}^2 = 0$. In our analysis, we will assume the former. This choice simplifies the equations and allows us to focus on the relevant aspects of the theory. It is worth noticing that ${\cal O}(\phi^4)$ contributions in the function $f(\phi)$ are finite when considering the on-shell action. Therefore, they do not play a role in the discussion.
 
Using the asymptotic expansion to solve the Einstein field equations order by order one finds that 
\begin{align}
    g_{(2)ij} ={}& -{\cal S}\left(g_{(0)}\right)-\frac{\kappa}{4}\left(1-\frac{48}{\ell^2}f_{(2)}\right)\phi_{(0)}^2g_{(0)ij}~, \nonumber \\ \nabla^i g_{(3)ij} ={}& \frac{2\kappa}{3\ell^2}\left\{\left[32\ftwo\phi_{(1)} - 8\kappa\ftwo\left(1-\frac{48}{\ell^2}\ftwo\right)\phi_{(0)}^2\right]\partial_j \phi_{(0)} - \left(\ell^2-32\ftwo\right)\phi_{(0)}\partial_j\phi_{(1)}\right\}~,\nonumber\\ {\rm Tr}~g_{(3)} ={}& \frac{4\kappa}{3}\left(1-\frac{48}{\ell^2}\ftwo\right)\left(\frac{4\kappa}{3\ell^2}\ftwo \phi_{(0)}^2 - \phi_{(1)}\right)\phi_{(0)}\,,
\end{align}
which recovers the coefficient of the minimally coupled scalar theory if $f_{(2)} = 0$. An arbitrary on-shell variation of the action alongside the asymptotic expansion of the fields yield
\begin{align}
    \delta S_{\rm sGB} ={}& \frac{1}{2\kappa}\int_{\partial {\cal M}}d^3x \sqrt{-\bg}\frac{1}{\rho^{3/2}}\Big[ \left(\ell^2 - 8\kappa f_{(0)} \right)\bg^{ij}\delta \bg_{ij}+\rho\Big\{\left(16\kappa f_{(0)} -2\ell^2\right)\bg^{ij}\delta\left(\partial_\rho \bg_{ij}\right) \nonumber \\ {}&-2\ell^2\kappa\bphi\delta\bphi  - \left(8\kappa f_{(0)}\left({\cal R}(\bg)^{ij}-\frac12 {\cal R}(\bg)\bg^{ij} + \bg^{ij}\bg^{mn}\partial_\rho \bg_{mn}\right)+24f_{(2)}\kappa\bphi^2\bg^{ij}\right. \nonumber \\ {}&+\ell^2\partial_\rho \bg^{ij}\Big)\delta\bg_{ij}\Big\} + {\cal O}\left(\rho^{2}\right)\Big]~.
\end{align}
If the zeroth-order coefficient of the scalar function expansion is chosen as
\begin{align}\label{f0}
    f_{(0)} = \frac{\ell^2}{8\kappa}~,
\end{align}
then, the Einstein-Hilbert sector becomes finite and a well-defined variational principle is achieved in terms of the sources without the need of a GHY term. This coupling coincides with the one obtained in Ref.~\cite{Olea:2005gb} for pure Einstein-AdS gravity. In four dimensions, the GB term is purely topological. This means that adding it to gravity action does not introduce modifications to the bulk dynamics, even though it changes the value of the on-shell action and conserved charges in a nontrivial way. Moreover, the Einstein-Hilbert action coupled to the GB term with the coupling \eqref{f0} on-shell is a sector of conformal gravity as shown in Ref.~\cite{Miskovic:2009bm}. The possibility to embed Einstein gravity in conformal gravity has shown to be useful in renormalizing gravity coupled to conformally coupled scalar fields~\cite{Anastasiou:2022wjq}.

The inclusion of the GB invariant together with the counterterm in Eq.~\eqref{Sphi} is sufficient to have a renormalized on-shell action. To see this, let us first define 
\begin{align}
    \bT_{\mu\nu} :={}& \left(T_{\mu\nu} - \frac12 g_{\mu\nu} T_{\mu\nu} \right) + \left(C_{\mu\nu} - \frac12 g_{\mu\nu}C\right)~.
\end{align}
This structure appears in the on-shell value of the Weyl tensor and the GB density. Using the on-shell relation
\begin{align}
    %R^2 - 4R^{\mu\nu}R_{\mu\nu} ={}& \kappa^2\left(\bT^2-4\bT^{\mu}_\nu \bT^\nu_\mu\right)~, \nonumber \\
W_{\alpha\beta}^{\mu\nu}W_{\mu\nu}^{\alpha\beta} ={}& R_{\alpha\beta}^{\mu\nu}R_{\mu\nu}^{\alpha\beta} - \frac{24}{\ell^2} + \frac{4\kappa}{\ell^2}\bT + \kappa^2\left(\delta^{\mu\nu}_{\alpha\beta}\delta^{\alpha\lambda}_{\mu\eta}\bT^{\beta}_{\lambda}\bT^{\eta}_{\nu} - 4\bT^{\mu}_\nu\bT^\nu_\mu - \frac23\bT^2\right)~,
\end{align}
we can write the GB density in terms of the square of the Weyl tensor. Then, the on-shell action can be written as
\begin{align}
    S_{\rm sGB} ={}& \int_{\cal M}d^4x\sqrt{-g}\Bigg[-\frac{3}{\kappa\ell^2} +\frac12\bT+ \left(f(\phi)-\frac12\phi f'(\phi)\right)\Bigg(W^{\mu\nu}_{\alpha\beta}W^{\alpha\beta}_{\mu\nu}+\frac{24}{\ell^2} \nonumber  \\ {}&-\frac{4\kappa}{\ell^2}\bT + \kappa^2\left(\frac53\bT^2 - \delta^{\mu\nu}_{\alpha\beta}\delta^{\alpha\lambda}_{\mu\eta}\bT^{\beta}_{\lambda}\bT^{\eta}_{\nu}\right)\Bigg)\Bigg] -\frac12\int_{\partial\cal M} d^3x \sqrt{-h}\,n^\mu\phi\partial_\mu\phi~.\end{align}
Notice that
\begin{align}
    f(\phi) - \frac12\phi f'(\phi) = \fzero -\frac32\fthree\phi^3-2\ffour\phi^4 + \dots = \sum_{n=0}^\infty \left(1-\frac{n}{2}\right)f_{(n)}\phi^n~,
\end{align}
does not contain quadratic terms. Then, the sGB coupling cannot be used to remove quadratic self-interaction terms.  

Since the Weyl square term is finite for AAdS spacetimes, we can choose the value of Eq.~\eqref{f0} such that the first two bulk terms cancel. Then, we are left only with quadratic terms in $\bT$. These terms remain finite if the action contains quadratic kinetic terms and/or quadratic self-interacting potentials as in the case in sGB gravity. Hence, the only divergences that cannot be eliminated by the GB density are those associated with the minimally coupled scalar field, that can be renormalized using the counterterm in Eq.~\eqref{Sphi}. As a result, the renormalized action can be expressed as 
\begin{align}
    S_{\rm sGB}^{\rm ren} = S_{\rm sGB} + S_{\phi}~,
\end{align}
where $f(\phi) = \frac{\ell^2}{8\kappa} + f_{(2)}\phi^2$, even though higher-order terms could be considered as they give finite contributions. 

The holographic 1-point for the scalar field depends on the boundary conditions and is controlled $W\left(\phi_{(0)}\right)$. For the holographic stress-energy tensor we obtain\footnote{For Einstein gravity with a minimally coupled scalar field, i.e., $f_{(2)} = 0$, the stress-energy tensor matches exactly that of Ref.~\cite{Caldarelli:2016nni}. Our notation relates to theirs by identifying ${\cal F}(\phi_{(0)}) = -\ell^2 W(\phi_{(0)})$.}
\begin{align}
    \langle T_{ij}\rangle ={}& -\ell^2\left[W\left(\phi_{(0)}\right) - \left( 1-\frac{32}{\ell^2}f_{(2)} \right)\phi_{(0)}\phi_{(1)} + \frac{8\kappa}{\ell^2}f_{(2)}\left(1-\frac{{48}}{\ell^2}f_{(2)}\right)\phi_{(0)}^3\right]g_{(0)ij} \nonumber \\ {}& + \frac{3\ell^2}{2\kappa}g_{(3)ij}~,
\end{align}
whose trace yields
\begin{align}
    \langle T\rangle =  -3\ell^2\left[W\left(\phi_{(0)}\right) -\frac13\phi_{(0)}\phi_{(1)}-\frac{16\kappa}{3\ell^2}f_{(2)}\left(1-\frac{{48}}{\ell^2}f_{(2)}\right)\phi_{(0)}^3\right]~.
\end{align}
Notice that if we consider mixed boundary conditions that respect conformal invariance, i.e., $W\left(\phi_{(0)}\right) = C\phi_{(0)}^3$, with $C$ some arbitrary constant (see Ref.~\cite{Henneaux:2006hk, Caldarelli:2016nni}), and using the fact that $W'\left(\phi_{(0)}\right)=\phi_{(1)}$ [cf. Eq.~\eqref{dWJ}] one obtains
\begin{align}
    \langle T\rangle = 16\kappa f_{(2)}\left(1-\frac{{48}}{\ell^2} f_{(2)}\right)\phi_{(0)}^3~.
\end{align}
Focusing on the nontrivial quadratic self-interaction, that is $\ftwo\neq0$, we find that conformal invariance of the boundary CFT implies\footnote{For more examples of breaking scaling symmetry due to the inclusion of scalar field, see Ref.~\cite{Caldarelli:2016nni}.}
\begin{align}\label{f2sol}
    f_{(2)} = \frac{\ell^2}{{48}}~.
\end{align}
Thus, the effective mass defined in Eq.~\eqref{meff} becomes $m_{\rm eff}^2\ell^2 = m^2\ell^2 - 1$. Then, replacing this value into Eq.~\eqref{BFmeff}, we find a modified BF bound for the bare mass $m$ triggered by the scalar-Gauss-Bonnet coupling, that is,
\begin{align}
    -\frac{5}{4}<m^2\ell^2\,.
\end{align}

The above results indicate that the addition of the GB term in the bulk is useful to renormalize the bulk theory, in a similar fashion as in the pure AdS gravity case. This invariant, when expressed as a boundary term, can be thought of as an extrinsic counterterm series. It is noteworthy that the boundary contribution of the GB term in four dimensions cannot be seen as a quasilocal stress tensor. However, it does contribute to the holographic stress tensor and renders the variation of the action finite. 

If we consider Dirichlet boundary conditions, denoted as $W\left(\phi_{(0)}\right) = 0$, the variation of the renormalized action introduces an additional piece arising from the boundary value of the scalar field. As a result, the vacuum expectation value of the boundary scalar operator can be expressed as 
\begin{align}
    \langle {\cal O}\rangle = \frac{1}{\sqrt{-g_{(0)}}}\frac{\delta S_{\rm H}^{\rm ren}}{\delta \phi_{(0)}} = \lim_{\epsilon \to 0}\left(\frac{1}{\epsilon^{\Delta-d}}\frac{1}{\sqrt{-h}}\frac{\delta S_{\rm H}^{\rm ren}}{\delta \phi}\right) = -\phi_{(1)} = -(2\Delta-d)\phi_{(2\Delta-d)}~,
\end{align}
which matches the holographic 1-point function of the scalar operator dual to a scalar field on an AdS background with Dirichlet boundary conditions.

If we take instead $\fthree \neq 0 $ and $\phi_{(0)}^2 = 0$~, we found that asymptotic analysis of the field equations gives
\begin{align}
    g_{(2)} ={}& -{\cal S}\left(g_{(0)}\right)_{ij}~, \nonumber \\ \nabla^i_{(0)}g_{(3)ij} ={}& \frac{2\kappa}{3\ell^2}\left[32\ftwo \phi_{(1)}\partial_j\phi_{(0)}-\left(\ell^2-32\ftwo\right)\phi_{(0)}\partial_j\phi_{(1)} 
 \right]~,\nonumber \\ {\rm Tr}~g_{(3)} ={}& -\frac{4\kappa}{3\ell^2}\left(\ell^2-48\ftwo\right)\phi_{(0)}\phi_{(1)}~.
\end{align}
Using the same counterterm as in the previous case, given in  Eq.~\eqref{Sphi}, we find that
\begin{align}
    \langle T_{ij}\rangle = \left(\ell^2-32\ftwo\right)\phi_{(0)}\phi_{(1)}g_{(0)ij}+\frac{3\kappa}{2\ell^2}g_{(3)ij}~,
\end{align}
which, independent of the boundary conditions, is traceless. Moreover, considering Dirichlet boundary conditions one gets
\begin{align}
    \langle {\cal O}\rangle = -\phi_{(1)} = -(2\Delta-d)\phi_{(2\Delta-d)}~,
\end{align}
just as in the previous scenario. 

Another interesting scenario to explore involves setting $\Delta = 3$. In this case, the scalar field behaves as
\begin{align}
    \phi(\rho,x^i) = \phi_{(0)}(x^i)+\rho\phi_{(2)}(x^i) + \rho^{3/2}\phi_{(3)}(x^i)+\dots~,
\end{align}
near the boundary. The relation between the mass and $\Delta$ becomes
\begin{align}
    m^2\ell^2 = \frac{48}{\ell^2}\ftwo~.
\end{align}

Solving the field equations order by order, we can derive several conditions on the scalar couplings $f_{(n)}$. First, the zeroth-order equations yield $\fone = 0 = \fthree$, or alternatively, $\fone = 0$ together with $\phi_{(0)}^2 = 0$. Assuming the former condition, the Einstein equations impose $\ftwo = 0$. Then, we obtain
\begin{align}
    g_{(2)ij} ={}& -{\cal S}(g_{(0)})_{ij} - \frac{\kappa}{4}g_{(0)ij}\partial^m \phi_{(0)}\partial_m \phi_{(0)} + \kappa \partial_i \phi_{(0)}\partial_j \phi_{(0)}~, \nonumber \\ \phi_{(2)} ={}& \frac12\Box_{(0)}\phi_{(0)}~, \nonumber  \\ {\rm Tr}~g_{(3)} ={}& 0~, \nonumber \\ \nabla^i_{(0)}g_{(3)ij} ={}& \frac{2\kappa}{\ell^2}\phi_{(1)}\partial_j\phi_{(0)}~.\label{g3dgbdelta3} 
\end{align}
Arbitrary variations of the on-shell action make evident that selecting the zeroth-order coupling in Eq.~\eqref{f0} eliminates the leading-order divergences. However, to address the remaining divergences associated with the scalar field, we need to introduce an appropriate counterterm. We have determined that including the intrinsic counterterm
\begin{align}
    S_{\rm sGB}^{\rm ct} = -\frac{\ell}{2}\int_{\partial\cal M}d^3x \sqrt{-h}h^{ij}\partial_i\phi\partial_j \phi~,
\end{align}
together with the choice in Eq.~\eqref{f0}, renders the renormalized action 
\begin{align}
    S_{\rm sGB}^{\rm ren} = S_{\rm sGB} + S_{\rm sGB}^{\rm ct}~,
\end{align}
finite on shell. This approach allows us to handle and regularize the divergences encountered in the theory. The resulting holographic stress tensor is given by
\begin{align}
    \langle T^{ij}\rangle = \frac{3\ell^2}{2\kappa}g_{(3)}^{ij}~.
\end{align}
Remarkably, this holographic stress-energy tensor is traceless as can be seen from Eq.~\eqref{g3dgbdelta3}. Additionally, we find a non-zero vacuum expectation value for the boundary scalar operator when we consider Dirichlet boundary conditions
\begin{align}
    \langle {\cal O}\rangle = -3\phi_{(3)} = -(2\Delta-d)\phi_{(2\Delta-d)}~.
\end{align}
This expectation value is akin to the behavior observed for the scalar field in AdS. These results provide insights into the behavior of bulk fields and their dual operators in the context of the AdS/CFT correspondence.

Finally, considering $\phi_{(0)}^2 = 0$ while keeping $\fthree$ unconstrained, the holographic data now reads
\begin{align}\label{last}
    g_{(2)ij} ={}& -{\cal S}(g_{(0)})_{ij} -\frac{\kappa}{4} g_{(0)ij}\partial_m\phi_{(0)}\partial^m\phi_{(0)}~,\nonumber \\ \phi_{(2)} ={}& \frac{1}{2}\left(1-\frac{72}{\ell^2}\fthree \phi_{(0)}\right)^{-1}\Box_{(0)} \phi_{(0)}~, \nonumber \\ \rm{Tr}~g_{(3)} ={}& \frac{64\kappa}{\ell^2}\ftwo\phi_{(0)}\phi_{(3)}~, \nonumber \\ \nabla^i_{(0)}g_{(3)ij} ={}& \frac{2\kappa}{3\ell^2}\left[\left(3+\frac{32}{\ell^2}\ftwo\right)\phi_{(3)}\partial_j\phi_{(0)}+\frac{32}{\ell^2}\ftwo\phi_{(0)}\partial_j\phi_{(3)}\right]~.
\end{align}
In this case, the holographic stress tensor is
\begin{align}
    \langle T_{ij}\rangle = \frac{3\ell^2}{2\kappa}g_{(3)ij}-\frac{32}{\ell^2}\ftwo \phi_{(0)}\phi_{(3)}~,
\end{align}
which is traceless as a consequence of Eq.\eqref{last}.

%%%%%%%%%%%%%%%%%%%%%

\section{Kinetic coupling to the Einstein tensor}\label{Sec:Horni}

A particular sector of the Horndeski theory, which has been widely studied, considers the nonminimal coupling of the scalar field to the Einstein tensor. This term belongs to the Horndeski class of gravity theories and it does not introduce higher-curvature terms. This coupling has been studied holographically to break the viscosity/entropy bound~\cite{Policastro:2001yc,Son:2002sd, Kovtun:2003wp} of ideal fluids without the need of adding higher-curvature corrections~\cite{Buchel:2003tz}. This serves to study how $\alpha'$ corrections to the hydrodynamics modify the holographic description of the boundary theory.

In the case under consideration here, a scalar coupling of the GB term is also included
\begin{align}
    S_{\rm H} = \int_{\cal M}d^4x \sqrt{-g}\left(\frac{R-2\Lambda}{2\kappa} - \frac12\left(\partial \phi\right)^2 - \frac12 m^2\phi^2  + f(\phi){\cal G} + \frac{\eta}{2} G_{\mu\nu}\nabla^\mu\phi\nabla^\nu\phi \right)~,
\end{align}
The field equations can be obtained by demanding arbitrary variations for the metric and the scalar field, giving
\begin{align}\label{EOMHorGB}
    G_{\mu\nu} + \Lambda g_{\mu\nu}  ={}& \kappa\left(T_{\mu\nu} + C_{\mu\nu} + H_{\mu\nu}\right)~, \\ \nabla_\mu\left[\left(g^{\mu\nu}-\eta G_{\mu\nu}\right)\nabla_\nu\phi\right] ={}& m^2\phi - f'(\phi){\cal G}~,
\end{align}
respectively, where $T_{\mu\nu}$ and $C_{\mu\nu}$ have been defined in Eqs.~\eqref{Tmunuminimalcoupling} and~\eqref{Cmunu}, and 
\begin{align}
    H_{\mu\nu} ={}& \frac{\eta}{4}\Big[ \delta^{\sigma\alpha\beta}_{\mu\lambda\rho}\nabla_\sigma\phi\nabla_\nu\phi R_{\alpha\beta}^{\lambda\rho}+\delta^{\rho\lambda\sigma}_{\alpha\beta\mu}\nabla_\rho\phi\nabla^{\alpha}\phi R^{\beta}_{~\nu\lambda\sigma} \nonumber \\ {}&+ 2g_{\mu\nu}G_{\alpha\beta}\nabla^{\alpha}\phi\nabla^\beta\phi +2\delta^{\alpha\beta\lambda}_{\rho\sigma\nu}g_{\lambda\mu}\nabla^\sigma\phi\nabla_\beta\phi\left(\nabla_\alpha\phi\nabla^\rho\phi\right) \Big]~, 
\end{align}
is the contribution to the field equations coming from the scalar-kinetic coupling to the Einstein tensor. 

For a massless scalar field, the absence of the GB coupling implies that this theory is endowed with a shift symmetry in field space, namely, the field equations remain invariant under $\phi\to\phi+\tilde{\phi}_0$, where $\tilde{\phi}_0$ is a constant. This is the typical symmetry exhibited by Galileons~\cite{Nicolis:2008in,Deffayet:2009wt,Deffayet:2009mn}.\footnote{If the scalar coupling to the GB is linear, the theory still has shift symmetry. However, this possibility was excluded by the dynamics as shown in the previous section.} In that case, however, there exists a no-hair theorem which prevents from finding black hole solutions with a nontrivial scalar field~\cite{Hui:2012qt}. Nevertheless, a suitable on-shell condition on the metric can be imposed such that the no-hair theorem is circumvented, allowing for asymptotically locally flat black holes~\cite{Rinaldi:2012vy}. Later, the same idea was extended to the case with the cosmological constant~\cite{Anabalon:2013oea} and with Maxwell fields~\cite{Cisterna:2014nua}. In this section, we consider the coupling between the scalar field and the GB term such that the shift symmetry is broken, to see how the BF bound is modified with respect to the one found in the previous section.  

\subsection{$\Delta = 2$}

Performing the asymptotic analysis in the presence of the kinetic coupling to the Einstein tensor, we find that the effective mass relation is modified according to 
\begin{align}\label{BFboundHorndeski}
    m_{\rm eff}^2\ell^2 = \Delta(\Delta -3)\left(1-\frac{3}{\ell^2}\eta\right) = -2\left(1-\frac{3}{\ell^2}\eta\right)\,,
\end{align}
where we have used $\Delta=2$ in the last equality.
% \rc{In this case, the BF bound turns out to be 
% \begin{align}
%     -\frac{\ell^2}{192}\left[9-4\Delta(3-\Delta)\left(1-\frac{3}{\ell^2}\eta\right)\right] < \ftwo < -\frac{\ell^2}{192} \left[5-4\Delta(\Delta-3)\left(1-\frac{3}{\ell^2}\eta\right)\right]\,,
% \end{align}
% and, focusing on the case $\Delta=2$, it becomes
% \begin{align}
%         -\frac{\ell^2}{192}\left(1+\frac{24}{\ell^2}\eta\right) < \ftwo < \frac{\ell^2}{64}\left(1-\frac{8}{\ell^2}\eta\right)~.
% \end{align}}
The boundary scalar equations impose that $\fone =0$~, together with either $\fthree = 0$ or $\phi_{(0)}^2 = 0$. We choose $\fthree = 0$ to ensure that the boundary value of the scalar field remains unconstrained, displaying a nontrivial interaction with the boundary metric. Then, solving the equations of motion order by order, the coefficients are found to be
\begin{align}
    g_{(2)ij} ={}& -{\cal S}(g_{(0)})_{ij} - \frac{\kappa}{4}\left(1-\frac{48}{\ell^2}\ftwo-\frac{5}{\ell^2}\eta\right)\phi_{(0)}^2g_{(0)ij}~, \nonumber \\
    {\rm Tr}~g_{(3)} ={}& \frac{4\kappa}{3}\left[\frac{4\kappa}{3\ell^2}\ftwo\left(1-\frac{5}{\ell^2}\eta - \frac{48}{\ell^2}\ftwo\right)\phi_{(0)}^3-\left(1-\frac{6}{\ell^2}\eta-\frac{48}{\ell^2}\ftwo\right)\phi_{(0)}\phi_{(1)}\right]~, \nonumber \\ \nabla^i_{(0)}g_{(3)ij} ={}& \frac{2\kappa}{3\ell^2}\Bigg\{\left[\left(32\ftwo+2\eta\right)\phi_{(1)}-8\kappa\ftwo\left(1-\frac{48}{\ell^2}\ftwo -\frac{5}{\ell^2}\eta\right)\phi_{(0)}^2 \right]\partial_j \phi_{(0)}\nonumber \\ {}& - \left(\ell^2-32\ftwo-5\eta\right)\phi_{(0)}\partial_j\phi_{(1)}\Bigg\}~.
\end{align}

Let us consider the variational problem. Using the asymptotic expansion and taking an arbitrary variation of the fields, we find
\begin{align}
    \delta S_{\rm H} ={}& \int_{\partial\cal M}d^3x \sqrt{-\bg}\rho^{-\frac32}\Big\{\left(\ell^2-8\kappa f_{(0)}\right)\bg^{ij}\left(\delta \bg_{ij}+2\rho\delta\partial_\rho \bg_{ij}\right) \nonumber \\ {}& -2\rho\Big[\left(\ell^2-3\eta\right)\bphi\delta\bphi- \left(4\kappa f_{(0)}\left(\frac12{\cal R}(\bg)\bg^{ij} + \bg^{ij}\bg^{mn}\partial_\rho\bg_{mn} - {\cal R}(\bg)_{mn}\bg^{im}\bg^{jn}\right)\right. \nonumber \\ {}& \left.\left.\left.- 2\kappa\left(\eta+12\ftwo\right)\bphi^2\bg^{ij} +\frac{\ell^2}{2}\bg^{im}\bg^{jn}\partial_\rho \bg_{mn}\right)\delta\bg_{ij}\right] + {\cal O}(\rho^2)\right\}~.
\end{align}
Notice that, in this case, Eq.~\eqref{f0} also removes the divergences coming from the gravitational sector. Additionally, we must take into account the counterterm in Eq.~\eqref{Sphi} associated with the scalar field. However, this counterterm should involve the kinetic coupling to the Einstein tensor. Then, the renormalized action turns out to be
\begin{align}\label{SrenHor}
    S_{\rm H}^{\rm ren} = S_{\rm H} + \frac{1}{\ell}\left(1-\frac{3}{\ell^2}\eta\right)S_{\phi}~.
\end{align}
Similar to the previous section, we can write the GB in terms of the square of the Weyl tensor and check that the on-shell action in Eq.~\eqref{SrenHor} is, indeed, finite.  For the holographic stress-energy tensor, we find that it is given by
\begin{align}
    \langle T_{ij} \rangle ={}& -\ell^2 g_{(0)ij}\Bigg[\left(1-\frac{3}{\ell^2}\eta\right)W\left(\phi_{(0)}\right)-\left(1-\frac{32}{\ell^2}f_{(2)}-\frac{5}{\ell^2}\eta\right)\phi_{(0)}\phi_{(1)} \nonumber \\ {}&+8\kappa f_{(2)}\left(1-\frac{5}{\ell^2}\eta-\frac{48}{\ell^2}f_{(2)}\right)\phi_{(0)}^3\Bigg]+ \frac{3\ell^2}{2\kappa}g_{(3)ij}~.
\end{align}
Imposing suitable boundary conditions, i.e., $W(\phi) = C \phi_{(0)}^3$~, the trace of the latter becomes
\begin{align}
    \langle T\rangle ={}& -16\kappa f_{(2)}\left(1-\frac{48}{\ell^2}f_{(2)}-\frac{5}{\ell^2}\eta\right)\phi_{(0)}^3~.
\end{align}
Therefore, conformal invariance is preserved in the boundary field theory if the trace vanishes. This implies that $f_{(2)} = 0$~, or
\begin{align}
    f_{(2)} = \frac{\ell^2}{48}\left(1-\frac{5}{\ell^2}\eta\right)~.
\end{align}
The latter condition yields an effective mass for the scalar field given by
\begin{align}\label{meffHorndeski}
    m_{\rm eff}^2\ell^2 = m^2\ell^2-1 + \frac{5}{\ell^2}\eta\,,
\end{align}
which is continuously connected with the value obtained in Sec.~\ref{Sec:SGB} in the limit $\eta\to0$. Then, the modified BF bound can be obtained by replacing Eq.~\eqref{meffHorndeski} into Eq.~\eqref{BFmeff}, giving
\begin{align}\label{BFboundHorndeski}
    -\frac{5}{4} < m^2\ell^2 + \frac{5}{\ell^2}\eta~.
\end{align}

Moreover, if we consider Dirichlet boundary conditions, i.e. $W\left(\phi_{(0)}\right) = 0$, we find an additional contribution to the vacuum expectation value of the boundary scalar coming from the kinetic coupling, namely,
\begin{align}
    \langle {\cal O}\rangle = -\left(1-\frac{3}{\ell^2}\eta\right)\phi_{(1)}=-\left(1-\frac{3}{\ell^2}\eta\right)(2\Delta-d)\phi_{(2\Delta-d)}~.
\end{align}
This modifies the result of the minimally coupled scalar field by a factor of $(1-3\eta\ell^{-2})$. Moreover, the diffeomorphism invariance of $g_{(0)}$ at the boundary implies a holographic Ward identity. While the coefficient in the expansion at a holographic order, $g_{(3)}$, can not be determined, we can determine its trace and divergence throughout the field equations. Then, the Ward identity reads
\begin{align}
    \nabla^i_{(0)}\langle T_{ij}\rangle = \left(1-\frac{3}{\ell^2}\eta\right)\phi_{(1)}\partial_j \phi_{(0)} = -\langle{\cal O}\rangle\partial_j\phi_{(0)}~,
\end{align}
which matches the result of Ref.~\cite{deHaro:2000vlm} for the minimally coupled scalar field. This computation of the holographic energy tensor-momentum tensor is not associated to a quasi-local stress tensor on the gravity side. However,  the coupling of matter renders its divergence different from zero, as it is related to the flow of momentum out of the boundary \cite{Brown:1992br}.

Let us examine Horndeski theory by considering a different possibility. For instance, if we fix $\Delta=2$ with $\phi_{(0)}^2 = 0$ while keeping $\fthree$ arbitrary, we obtain
\begin{align}
    g_{(2)ij} ={}& -{\cal S}(g_{(0)})_{ij}~, \nonumber \\ {\rm Tr}~ g_{(3)} ={}& -\frac{4\kappa}{3}\left(1-\frac{48}{\ell^2}\ftwo - \frac{6}{\ell^2}\eta\right)\phi_{(0)}\phi_{(1)}~, \nonumber \\ \nabla_{(0)}^i g_{(3)ij} ={}& \frac{2\kappa}{3\ell^2}\left[\left(32\ftwo+2\eta\right)\phi_{(1)}\partial_j\phi_{(0)} - \left(\ell^2-32\ftwo-5\eta\right)\phi_{(0)}\partial_j\phi_{(1)} \right]~.
\end{align}
The variation of the on-shell action gives the same kind of divergences as in the previous case. Therefore, the renormalized action must be the one in Eq.~\eqref{SrenHor}. The holographic stress tensor in this case becomes
\begin{align}
    \langle T_{ij}\rangle = \ell^2\left(1-\frac{32}{\ell^2}\ftwo - \frac{5}{\ell^2}\eta\right)\phi_{(0)}\phi_{(1)}g_{(0)ij} + \frac{3\ell^2}{2\kappa}g_{(3)ij}~,
\end{align}
which is always traceless as $\phi_{(1)}$ is proportional to positive powers of $\phi_{(0)}$ when considering mixed boundary conditions. Therefore, considering $\phi_{(0)}$ to be infinitesimal, one has a well-defined boundary CFT with a continuous $\eta \to 0$ limit. Finally, using Dirichlet boundary conditions we find that the vacuum expectation value and the Ward identity associated to the boundary scalar operator are the same as in the case with $\fthree = 0$ and $\phi_{(0)}$ unfixed. 
\subsection{$\Delta = 3$}
Consider now the case $\Delta = 3$. For this choice, the mass simply becomes $m^2 \ell^4 = 48\ftwo$ and the scalar equation imposes either $\fone = 0 =\fthree$ or $\fone = 0 = \phi_{(0)}^2$. Moreover, the zeroth order of the Einstein equations restricts further the theory with $\ftwo = 0$ or $\phi_{(0)}^2 = 0$~. Focusing first on the case when $\phi_{(0)}$ is arbitrary and solving for the coefficients of the metric, we obtain
\begin{align}
    \gtwoij ={}& -{\cal S}(g_{(0)})_{ij} - \frac{\kappa}{4}\left(1-\frac{3}{\ell^2}\eta\right)\gzij \partial_m\phi_{(0)}\partial^m\phi_{(0)} + \kappa\left(1-\frac{2}{\ell^2}\eta\right)\partial_i\phi_{(0)}\partial_j \phi_{(0)}~,\nonumber \\ \phi_{(2)}={}& \frac12 \Box_{(0)}\phi_{(0)} ~, \nonumber \\ {\rm Tr}~g_{(3)} ={}& 0\,, \nonumber \\ \nabla^i_{(0)}g_{(3)ij} ={}& 2\kappa\left(1-\frac{3}{\ell^2}\eta\right)\phi_{(3)}\partial_j\phi_{(0)} ~.
\end{align}
An arbitrary on-shell variation of the action with $\fzero = \ell^2/8\kappa$ yields
\begin{align}
    \delta S_{\rm H} = {}&-\frac{1}{2}\int_{\partial\cal M}d^3x \sqrt{-\bg}\rho^{-1/2}\Big[\kappa\eta\left(\bg^{mi}\bg^{nj}\partial_i\bphi\partial_j\bphi \delta \bg_{mn}-\bg^{ij}\partial_{m}\bphi\partial^m\bphi \delta \bg_{ij}\right)\nonumber \\ {}&+\ell^2\left({\cal R}(\bg)^{ij} - \frac12 {\cal R}(\bg)\bg^{ij}+\bg^{im}\bg^{jn}\partial_\rho \bg_{mn}\right)\delta\bg_{ij}+4\kappa\left(\ell^2-3\eta\right)\partial_\rho\bphi\delta\bphi \Big]~.
\end{align}
In order to eliminate the divergencies in this case, we notice that the same counterterm as in the sGB theory renders the theory finite but with a different coupling, that is,
\begin{align}\label{HorCt}
   S_{\rm H}^{\rm ct} = -\frac{\ell}{2}\left(1-\frac{3}{\ell^2}\eta\right)\int_{\partial \cal M}d^3x\sqrt{-h}h^{ij}\partial_i \phi \partial_j\phi~.
\end{align}
Then, the action
\begin{align}
    S_{\rm H}^{\rm ren} = S_{\rm H} + S_{\rm H}^{\rm ct}~,
\end{align}
is on-shell finite. Although the counterterm in Eq.~\eqref{HorCt} is necessary, most of the aforementioned solutions have a scalar profile that depends on the radial coordinate only, making the counterterm identically zero. Thus, the particular value of Eq.~\eqref{f0} is enough to remove the divergences of the theory. Nevertheless, it has been found that the theory admits regular solutions if one considers time-dependent scalar field~\cite{Babichev:2013cya}, such that the scalar does not inherit the spacetime symmetries, but the stress-tensor does. Moreover, there are interesting solutions of scalar-tensor gravity theories containing scalar fields that depend on both the radial and boundary coordinates such as accelerating black holes~\cite{Lu:2014ida, Cisterna:2021xxq, Cisterna:2023qhh, Barrientos:2023tqb}, whose holographic properties remain to be fully understood, and instantons~\cite{deHaro:2006wy,deHaro:2000vlm}, that can be used to explore vacuum decay of the boundary conformal theory~\cite{Papadimitriou:2007sj}. In this case, the holographic stress-energy tensor becomes
\begin{align}
    \langle T_{ij}\rangle = \frac{3\ell^2}{2\kappa}g_{(3)}^{ij}~,
\end{align}
which is traceless. Moreover, the vacuum expectation value of the boundary source becomes
\begin{align}
    \langle {\cal O}\rangle = -3\left(1 - \frac{3}{\ell^2}\eta\right)\phi_{(3)} = -\left(1 - \frac{3}{\ell^2}\eta\right)(2\Delta-d)\phi_{(2\Delta-d)}~.
\end{align}
This shows how the couplings of the theory modify the value of the 1-point functions of the dual operators. Moreover, the holographic Ward identity becomes
\begin{align}
    \nabla^i_{(0)}\langle T_{ij}\rangle = 3\left(1-\frac{3}{\ell^2}\eta\right)\phi_{(3)}\partial_j \phi_{(0)} = -\langle{\cal O}\rangle\partial_j \phi_{(0)}~,
\end{align}
just as in the previous cases. 

Finally, for an arbitrary $\ftwo$ and setting the source of the scalar operator to be infinitesimal, i.e. $\phi_{(0)}^2 = 0$, we find that the coefficients can be solved as
\begin{align}\label{g3hornidelta3}
    g_{(2)ij} ={}& -{\cal S}_{(0)ij}+ \kappa\left(1-\frac{2}{\ell^2}\eta\right)\partial_i\phi_{(0)}\partial_j\phi_{(0)}\nonumber \\ {}& + \frac{\kappa}{\ell^2}\left(24\ftwo \phi_{(0)}\phi_{(2)}-\frac14\left(\ell^2-3\eta\right)\partial^m\phi_{(0)}\partial_m \phi_{(0)}\right)g_{(0)ij} \nonumber \\ \phi_{(2)} = {}&\frac12\left(\frac{\ell^2-3\eta}{\ell^2-3\eta-72\ftwo\phi_{(0)}}\right)\Box_{(0)}\phi_{(0)} \nonumber \\ {\rm Tr}~g_{(3)} ={}& \frac{64\kappa}{\ell^2}\ftwo\phi_{(0)}\phi_{(3)}~, \nonumber \\ \nabla^i_{(0)}g_{(3)ij} ={}& \frac{2\kappa}{3\ell^2}\left[\left(3 + \frac{32}{\ell^2}\ftwo - \frac{9}{\ell^2}\eta\right)\phi_{(3)}\partial_j\phi_{(0)} + \frac{32}{\ell^2}\ftwo\phi_{(0)}\partial_j\phi_{(3)}\right]~,
\end{align}
Choosing $\fzero$ as in Eq.~\eqref{f0} and considering the same counterterm as in Eq.~\eqref{HorCt}, the on-shell action becomes finite and the variational principle is well-posed. Then, the holographic stress-energy tensor becomes
\begin{align}
    \langle T_{ij}\rangle = -\frac{32}{\ell^2}\ftwo\phi_{(0)}\phi_{(3)}g_{(0)ij} + \frac{3\ell^2}{2\kappa}g_{(3)ij}~,
\end{align}
which is traceless by virtue of Eq.~\eqref{g3hornidelta3}. If we consider Dirichlet boundary conditions, the scalar operator acquires a nontrivial vacuum expectation value, that is,
\begin{align}
    \langle {\cal O}\rangle = -3\left(1-\frac{3}{\ell^2}\eta\right)\phi_{(3)} = -\left(2\Delta-d\right)\left(1-\frac{3}{\ell^2}\eta\right)\phi_{(2\Delta-d)}~,
\end{align}
and a Ward identity which is given by
\begin{align}
    \nabla^i_{(0)}\langle T_{ij}\rangle = 3\left(1-\frac{3}{\ell^2}\eta\right)\phi_{(3)}\partial_j\phi_{(0)} = -\langle{\cal O}\rangle\partial_j\phi_{(0)}~,
\end{align}
just as in the case with an arbitrary $\phi_{(0)}$.

\subsection{Minimal Horndeski theory}\label{Sec:MinHorni}
Finally, we discuss on the particular case $f_{(n)} = 0 = m^2$~, $\forall n\in {\mathbb N}_{>0}$~\cite{Charmousis:2011ea, Charmousis:2011bf}, which has received a lot of attention in a holographic context~\cite{Filios:2018xvy, Kuang:2016edj, Li:2018kqp, Liu:2018hzo,Feng:2018sqm,Li:2018rgn,Jiang:2017imk} as it contains analytic solutions~\cite{Arratia:2020hoy, Anabalon:2013oea, Cisterna:2014nua, Cisterna:2015yla,Cisterna:2016vdx, Brihaye:2016lin, Babichev:2013cya, Babichev:2016rlq, Cisterna:2017jmv, Stetsko:2018fzt, Cisterna:2018mww}. The action in this case is
\begin{align}\label{SHormin}
    S_{\rm Hmin} = \int_{\cal M}d^4x\sqrt{-g}\left(\frac{R-2\Lambda}{2\kappa} - \frac12\left(\alpha g_{\mu\nu} - \eta G_{\mu\nu}\right)\nabla^\mu\phi\nabla^\nu\phi + f_{(0)}{\cal G}\right)~, 
\end{align}
where we have included an arbitrary constant for the canonical kinetic term; this is achieved by rescaling the scalar field and the metric.\footnote{Consider $\phi\to\alpha^{-1}\phi$ and $g_{\mu\nu}\to\alpha g_{\mu\nu}$, such that $R\to \alpha^{-1}R$. The Einstein tensor is scale-invariant, i.e., $G_{\mu\nu}\to G_{\mu\nu}$ and, rescaling, $\Lambda\to \alpha \Lambda~, ~\kappa \to \alpha \kappa$~, the minimal Horndeski theory with canonical kinetic term, i.e., $\alpha = 1$~, becomes that of Eq.~\eqref{SHormin}.} Additionally, we have included the GB term as it will serve as a counterterm in the scalar-tensor theories we are interested in. Recall that this sector of Horndeski theory is endowed with a shift symmetry as the scalar field appears only through derivatives in the action.

The field equations correspond to a subset of Eq.~\eqref{EOMHorGB}; explicitly, they are 
\begin{align}
G_{\mu\nu} + \Lambda g_{\mu\nu} ={}& \kappa\left( T_{\mu\nu}^{\rm min} + H_{\mu\nu}\right)~,\\ \nabla^\mu\left[\left(\alpha g_{\mu\nu}-\eta G_{\mu\nu}\right)\nabla^\nu\phi\right] = {}&0~,
\end{align}
where $T_{\mu\nu}^{\rm min} = \tfrac{\alpha}{2}T_{\mu\nu}$ with the latter defined in Eq.~\eqref{Tmunuminimalcoupling} by setting $m=0$.
Using the FG expansion, we find that the scalar field equation becomes
\begin{align}
   \Delta\left(\Delta-3\right)\left(\ell^2\alpha-3\eta\right)\bphi + {\cal O}\left(\rho\right)= 0~,
\end{align}
which is satisfied either if $\Delta = 3$ or $\eta = \frac{1}{3}\ell^2\alpha$ to all orders without fixing the conformal weight. If the latter point in the parameter space is assumed, then the action is renormalized simply by the GB term if one fixes $\fzero$ as in Eq.~\eqref{f0}. Nevertheless, this is a critical point of the theory. As shown in Refs.~\cite{Jiang:2017imk, Feng:2018sqm, Li:2018kqp}, the theory admits a solution that is nearly AdS with a nontrivial scalar field whose integration constant appears in the same footing as the cosmological constant. Therefore, to obtain a scalar field with a more interesting backreaction, in this Subsection, we focus on the $\Delta=3$ case with $\alpha\neq 3\eta\ell^{-2}$. This choice fixes Dirichlet boundary conditions for the scalar field; neither Neumann nor mixed boundary conditions are allowed for $\Delta=3$. Additionally, the scalar field becomes massless (see Eq.~\eqref{BFboundHorndeski} with $f_{(n)}=0$, $\forall n\in {\mathbb N}_{>0}$), it lies within the BF bound, and its holografic stress-energy tensor is traceless. This implies that there are no restrictions whatsoever on the parameter space coming from the compatibility of the BF bound, the conformal invariance at the boundary, and the absence of ghosts in the bulk theory---see Eq.~\eqref{boundkinetic} below.

In an Einstein-AdS background, the scalar sector of the minimal Horndeski theory becomes simply a minimally coupled massless scalar, that is,
\begin{align}
    {\cal L}_\phi = - \frac12\left(\alpha -\frac{3}{\ell^2}\eta\right)\left(\partial\phi\right)^2~.
\end{align}
Thus, the absence of ghosts implies the inequality
\begin{align}\label{boundkinetic}
    \alpha -\frac{3}{\ell^2}\eta \geq 0\,.
\end{align}
The bound is saturated at the critical point in which the scalar contribution vanishes. These kinds of critical points were also studied in the pure GB gravity \cite{Fan:2016zfs} and the black hole solution of Ref.~\cite{Anabalon:2013oea} simply becomes the Schwarzschild-AdS black hole with a vanishing scalar field. In this case, the holographic stress-energy tensor is traceless and it equals that of the pure gravity case. Nonetheless, it is possible to obtain an exact global AdS background with a nontrivial scalar. The latter breaks the AdS isometries and it contains a logarithmic mode in the FG expansion. This indicates that the dual theory is scale invariant but not conformally invariant (see Ref.~\cite{Nakayama:2013is} for details). 

Solving Einstein's equations order by order, we obtain $\fone = 0 = \fthree$ as before, and the coefficients of the metric expansion are now solved as
\begin{align}
    g_{(2)ij} ={}& -{\cal S}_{ij}\left(g_{(0)}\right) -\frac{\kappa}{4\ell^2}\left(\ell^2\alpha-3\eta\right)g_{(0)ij}\partial_{m}\phi_{(0)}\partial^m\phi_{(0)}\nonumber \\ {}&+ \frac{\kappa}{\ell^2}\left(\ell^2\alpha-2 \eta\right)\partial_i\phi_{(0)}\partial_j\phi_{(0)}~, \nonumber \\ {\rm Tr}~g_{(3)} ={}& 0~, \nonumber \\ \nabla^i_{(0)}g_{(3)ij} ={}& \frac{2\kappa}{\ell^2} \left(\alpha -\frac{3}{\ell^2}\eta\right)\phi_{(3)}\partial_j\phi_{(0)}~.
\end{align}
Considering an on-shell variation of the action with the GB coupling found in Eq.~\eqref{f0}, we arrive at
\begin{align}
    \delta S_{\rm Hmin} ={}& \frac12\int_{\partial\cal M}d^3x\sqrt{-g_{(0)}}\left[\rho^{-\frac12}\left(\ell^2\alpha-3\eta\right)\left(\frac12g_{(0)}^{ij}\phi_{(0)}^2-\partial^i\phi_{(0)}\partial^j\phi_{(0)}\right)\right. \nonumber \\ {}&\left.-\frac{3\ell^2}{2\kappa}g^{ij}_{(3)}\right]\delta g_{(0)ij}~.
\end{align}
This is finite if we choose the value in Eq.~\eqref{f0} but the critical value of $\eta$ is still needed. Nonetheless, the latter can be rendered arbitrary if we add a suitable counterterm, that is,
\begin{align}
    S_{\rm ct} = -\frac{1}{2\ell}\left(\ell^2\alpha-3\eta \right)\int_{\partial \cal M}d^3x \sqrt{-h}h^{ij}\partial_i \phi\partial_j\phi~.
\end{align}
With this counterterm, the action is finite. Then, the holographic stress-energy tensor is given by
\begin{align}
    \langle T_{ij}\rangle = \frac{3\ell^2}{2\kappa}g_{(3)ij}~,
\end{align}
and the vacuum expectation value for the boundary scalar operator and the holographic Ward identity become
\begin{align}
    \langle {\cal O}\rangle = -3\left(\alpha-\frac{3}{\ell^2}\eta\right)\phi_{(3)}=-\left(\alpha-\frac{3}{\ell^2}\eta\right)(2\Delta-d)\phi_{(2\Delta-d)}~,
\end{align}
and 
\begin{align}
    \nabla^i\langle T_{ij}\rangle = 3\left(\alpha - \frac{3}{\ell^2}\eta\right)\phi_{(3)}\partial_j\phi_{(0)} = -\langle {\cal O}\rangle \partial_{j} \phi_{(0)}~,
\end{align}
respectively. Notice that, at the critical point $\alpha = 3\eta\ell^{-2}$, the vacuum expectation value of the scalar field vanishes and the holographic stress-energy tensor becomes covariantly conserved. In this case, the known scalar-field solutions either vanish or do not backreact and the action becomes simply the one of Einstein's gravity. This shows that taking the limit to this critical value is consistent also at the level of 1-point functions. 
%%%%%%%%%%%%%%%%%%%%%%%%%%%%%%%
\section{Discussion}\label{Sec:Concs}

In this work, we have analyzed the holographic properties of different gravity theories of the Horndeski class. The aspects of the renormalization of both the action and its variation, for AAdS spaces, are worked out in great detail. To this end, we introduced the GB term nonminimally coupled to an arbitrary function of the scalar field. One of the main results is that the asymptotic analysis of the field equations restricts considerably the form of such a function. In particular, we found that the linear-scalar coupling to the GB term is not allowed in AAdS spaces as a consequence of the field equations. The sGB coupling contributes to an effective mass of the scalar field and we find how the BF bound is modified by the presence of this term in the bulk.

If the scalar-kinetic coupling to the Einstein tensor is introduced to the sGB theory, we find that its contribution shifts the BF bound. We have analyzed different possibilities and, in all cases, we obtained the counterterms that render the theory finite, the vacuum expectation value of the scalar operator at the boundary, and the holographic stress-energy tensor. Furthermore, we have studied the holographic Ward identity associated with coordinate transformations at the boundary and identified how the nonminimal coupling of Horndeski's theory produces the correct anomalous term. 
When considering mixed boundary conditions, we found that in all theories with $\Delta=3$ the deformations are marginal and there is no breaking of scale invariance at the boundary. However, in the case of $\Delta=2$~, we get that there is a possible conformal anomaly that depends on the scalar potential. In the case of the general coupling between the scalar field and GB invariant, the quadratic scalar induces on-shell a modification of the effective mass of the scalar. Therefore, for certain values of the $f_{(2)}$ coupling constant, there is a breaking of scaling symmetry producing an anomaly term in the dilatation Ward identity. Such anomalous terms can be also avoided by coupling the Einstein tensor with the kinetic term of the scalar and fixing the coupling in terms of the sGB potential.

The minimal Horndeski gravity theory considered in Eq.~\eqref{SHormin} encompasses numerous black hole solutions. However, only a limited subset among them possesses exact solutions for the scalar field, as usually only its radial derivative is known, rather than the full analytic expression. Nonetheless, there are a few cases in which this issue can be circumvented. In these situations, the solutions feature either negligible backreaction or modify the geometry as an effective cosmological constant. For instance, in Ref.~\cite{Anabalon:2013oea}, an analytic solution corresponding to a topological Schwarzschild black hole with a flat transverse section and such an effective cosmological constant was found. There exists an analytic solution for the scalar field, even though there are new logarithmic divergences near the boundary. For those cases, the additional divergence introduced in the action~\eqref{SHormin} can be thought of as coming from an effective cosmological, which appears as modified by the scalar field, such that can be properly renormalized. This approach requires a fine-tuned coupling for the GB term as in Eq.~\eqref{f0} but in terms of an effective cosmological constant. As a result, introducing the GB term with a different coupling is sufficient to renormalize the additional divergences in the action, even in the presence of logarithmic modes of the scalar field. 

Interesting questions remain open. In particular, extending this analysis to include the logarithmic modes is certainly very important, since some of the known analytic solutions in the literature are endowed with this asymptotic behavior. The latter will allow one to check explicitly the compatibility between the Noether-Wald entropy and the one obtained via Euclidean methods. Indeed, the additional boundary terms found here might contribute to the renormalized Euclidean on-shell action at finite order, modifying the thermodynamic variables. We expect that the latter will solve the discrepancy while satisfying the first law of thermodynamics. However, since the black hole solution of Ref.~\cite{Anabalon:2013oea} is endowed with logarithmic scalar modes, we postpone a deeper study of this point for a forthcoming paper. Additionally, it is well known that in even boundary dimensions, the holographic trace anomaly is related to the logarithmic modes of the metric~\cite{Henningson:1998gx}. The role of nonminimally coupled scalar fields in the holographic trace anomaly is indeed worth studying, alongside the generalization of the counterterms found here to higher dimensions. On the other hand, the results found here are useful for studying holographic measurements such as entanglement entropy and superconductivity, which may be the subject of future work.

\acknowledgments
We thank Giorgos Anastasiou, Ignacio J. Araya, Adolfo Cisterna, Jos\'e Figueroa, Julio Oliva, and Leonardo Sanhueza for helpful discussions. This work is partially funded by Agencia Nacional de Investigaci\'on y Desarrollo {\sc (ANID)} through Anillo Grant ACT210100 \emph{Holography and its applications to High Energy Physics, Quantum Gravity and Condensed Matter Systems}. The work of {\sc NC} is supported by {\sc Beca Doctorado Nacional} {\sc ANID} 2023 No. 21231876. The work of {\sc CC} is partially supported by {\sc ANID} through {\sc FONDECYT} Regular grants No 1240043, 1240048, 1230112, and 1210500. {\sc FD} is supported by {\sc Beca Doctorado nacional} 
({\sc ANID}) 2021 No. 21211335, and {\sc FONDECYT} Regular grant No. 1210500. The work of {\sc RO} has been funded by the {\sc FONDECYT} Regular Grants 1230492 and 1231779.

\begin{appendix}

\section{Scalar field boundary conditions}\label{App:BdryConditions}
%%%%%%%%%%%%%%%%%%%%%%%%%%%%

In order to stress the importance of boundary conditions in the dual field theory, let us analyze the case of a massive scalar field on global AdS as an example. The corresponding action can be expressed as
\begin{align}
    S = \frac12\int_{{\cal M}}d^{d+1}x\sqrt{g}\left[\left(\partial \phi\right)^2 + m^2 \phi^2\right]~,
\end{align}
where ${\cal M}$ is an Euclidean AdS$_{d+1}$ background with spacetime coordinates $x^\mu = (z,x^i)$. Its line element in the Poincar\'e patch is given by 
\begin{align}
    ds^2 = \frac{\ell^2}{z^2}\left(dz^2 + \delta_{ij}dx^i dx^j\right)~.
\end{align}
An arbitrary variation of the on-shell action with respect to the dynamic field yields
\begin{align}
    \delta S ={}& -\int_{\cal M} d^{d+1}x \sqrt{g}\left(\nabla^2 -m^2\right)\phi\delta\phi-\int_{\partial \cal M}d^d x \sqrt{h}\left( n^\mu \nabla_\mu\phi\right)\delta\phi\nonumber \\ ={}&- \int_{\partial\cal M}d^d x\left(\frac{z}{\ell}\right)^{1-d}\partial_z\phi \,\delta\phi ~.
\end{align}
Depending on the boundary conditions, an extra boundary term must be added such that the on-shell action possesses a minimum. For Dirichlet boundary conditions, the variation of the scalar field vanishes at the boundary, so there is no need to add such a term. For Neumann boundary conditions, however, the normal derivative of the scalar field is fixed. Thus, one needs to add 
\begin{align}
    S_{\rm N} = \int d^{d}x \left(\frac{z}{\ell}\right)^{1-d}\phi\partial_z \phi~,
\end{align}
to the bulk action, such that the variational principle is well-posed. Then, the source of the boundary scalar operator is given by the normal derivative of the field. This corresponds to the radial canonical momentum associated with the scalar field. Additionally, it is possible to impose mixed boundary conditions that involve a relationship between the scalar field and its normal derivative. The latter specifies the behavior of the boundary scalar operator according to
\begin{align}
    \psi := \phi+\lambda n^\mu\partial_\mu\phi~,
\end{align}
where $\lambda$ is a non-zero real number. Then, one needs to consider a boundary term of the form
\begin{align}
    S_{\rm M} = -\frac12 \int d^dx~ \left(\frac{z}{\ell}\right)^{1-d}\psi \partial_z \phi~,
\end{align}
and the source is now related to $\psi$~. 

On the Euclidean AdS$_{d+1}$ background, the field equation for the scalar field can be written as
\begin{align}
    z^{d+1}\partial_z\left(z^{1-d}\partial_z\phi\right) + z^2 \delta^{ij}\partial_i\partial_j\phi = m^2\ell^2 \phi~.
\end{align}
Near the boundary, one can check that the solution is,\footnote{If $\Delta =d/2$, one needs to consider solutions with logarithmic behavior.} 
\begin{align}
    \phi(z,x^i) \sim z^{d-\Delta}\phi_{(0)}(x^i) + z^\Delta\phi_{(1)}(x^i)~,
\end{align}
where
\begin{align}\label{AppBF}
    \Delta = \frac{d}{2}\pm\sqrt{\frac{d^2}{4}+m^2\ell^2}~.
\end{align}
In Ref.~\cite{Witten:1998qj}, it was shown that $\Delta$ corresponds to the conformal weight of a scalar operator of a $d$-dimensional CFT. This connection implies that for the scaling weight to be a real quantity, the condition $m^2\ell^2 > -d^2/4$ must be satisfied. Remarkably, this condition is compatible with the BF bound~\cite{Breitenlohner:1982bm}, suggesting that a scalar field in AdS can possess a negative mass while being stable. 
The functions $\phi_{(0)}$ and $\phi_{(1)}$ are the two linearly independent solutions of the second-order field equations. The leading-order term can be either singular if $\Delta > d$, trivial if $\Delta < d$, or constant if $\Delta = d$. Then, depending on the mass of the scalar field, the bulk geometry is modified while preserving the asymptotic structure. 

As it can be seen from Eq.~\eqref{AppBF}, the conformal weight $\Delta$ is a positive real number such that $\Delta > d-\Delta$. Therefore, $\phi_{(0)}$ is associated with non-normalizable modes at the boundary. To identify the source of the boundary scalar operator, one needs to consider
\begin{align}
    \varphi(x^i) = \lim_{z\to0}z^{\Delta-d}\phi(z,x^i)~,
\end{align}
which is always finite. Boundary conditions fix a function of $\phi_{(0)}$ and $\phi_{(1)}$ at the boundary, that corresponds to the source of the dual scalar operator. The relation between the modes reduces the degrees of freedom in the dual theory by one-half. For instance, Dirichlet boundary conditions fix the source to be $\phi_{(0)}$ and the remaining degree of freedom corresponds to the normalizable mode. As shown in Ref.~\cite{Papadimitriou:2004ap}, the non-normalizable mode $\phi_{(1)}$ does not transform properly under Weyl rescalings. Therefore, from a holographic viewpoint, one needs to consider the renormalized radial momentum $\hat\pi_\phi$, i.e., the first regular coefficient in the asymptotic expansion of the canonical momentum $\pi_\phi$. The latter usually differ from $\phi_{(1)}$ by a local functional of $\phi_{(0)}$. Then, for Neumann boundary conditions, one fixes $\hat\pi_\phi$ rather than $\phi_{(1)}$. This shows that the leading and sub-leading coefficients in the asymptotic expansion are canonically conjugated variables (see Ref.~\cite{Papadimitriou:2007sj} details). 

In general, one introduces a boundary term that depends on the scalar field and derivatives thereof, whose explicit form depends on the boundary conditions. As shown before, the field equations fix the relation between derivatives of both the field and the boundary term via boundary conditions. The extra boundary contribution to the gravity action implies a modification of the boundary theory. In the case of mixed boundary conditions, this corresponds to modifying the holographic CFT by multi-trace operators if the deformation function is built not only by the fields but also by the operators~\cite{Witten:2001ua}. Then, the vacuum expectation value of the dual operator with conformal dimension $d-\Delta$ corresponds to $\phi_{(0)}$ and its current related to $\phi_{(1)}$. Since the on-shell action is identified with the generating functional of connected correlators of the dual CFT, say $\Gamma$, the addition of the new term modifies the dual theory as
\begin{align}
    \Gamma[\phi_{(0)}] \to \Gamma[J] - \int d^dx \sqrt{g_{(0)}}J\phi_{(0)}~,
\end{align}
where $J$ is the current which depends on $\phi_{(0)}$ and $\hat\pi_\phi$ and is fixed by boundary condition in the string theory side. Following Ref.~\cite{Anabalon:2015xvl}, we have encoded the deformations of the boundary theory in $W\left(\phi_{(0)}\right)$, which must be fixed such that the variational principle is consistent with the corresponding boundary conditions. Then, the boundary CFT is deformed and the boundary conditions impose  \cite{Witten:2001ua}
\begin{align}
    J = \frac{dW\left(\phi_{(0)}\right)}{d\phi_{(0)}}~.
\end{align}
Using Neumann or mixed boundary conditions corresponds to modifying the conformal vacuum at the boundary. In the case of mixed boundary conditions, the multi-trace deformations could break the conformal invariance as they modify the $n$-point functions (see Ref.~\cite{Minces:1999eg}). Then, the deformations could be marginal, relevant, or irrelevant depending on the mass of the bulk scalar field. Moreover, multi-trace deformations have been associated to multi-particle states in the dual gravity theory~\cite{Aharony:2001pa}.

\end{appendix}

\bibliographystyle{JHEP}
\bibliography{biblio}

\end{document}